\documentclass[twocolumn,aps,superscriptaddress,amsmath,amssymb]{revtex4}
\usepackage{amsfonts}
\usepackage{bbm}

\usepackage{graphicx}
\usepackage{dcolumn}
\usepackage{bm}

\begin{document}
\title{Charge transport in a multi-terminal DNA tetrahedron: Interplay among contact position, disorder, and base-pair mismatch}

\author{Pei-Jia Hu}
\affiliation{Hunan Key Laboratory for Super-microstructure and Ultrafast Process, School of Physics and Electronics, Central South University, Changsha 410083, China}

\author{Si-Xian Wang}
\affiliation{Hunan Key Laboratory for Super-microstructure and Ultrafast Process, School of Physics and Electronics, Central South University, Changsha 410083, China}

\author{Xiao-Feng Chen}
\affiliation{Hunan Key Laboratory for Super-microstructure and Ultrafast Process, School of Physics and Electronics, Central South University, Changsha 410083, China}
\affiliation{School of Physical Science and Technology, Lanzhou University, Lanzhou 730000, China}

\author{Xiao-Hui Gao}
\affiliation{Hunan Key Laboratory for Super-microstructure and Ultrafast Process, School of Physics and Electronics, Central South University, Changsha 410083, China}

\author{Tie-Feng Fang}
\affiliation{School of Physical Science and Technology, Lanzhou University, Lanzhou 730000, China}

\author{Ai-Min Guo}
\email[]{aimin.guo@csu.edu.cn}
\affiliation{Hunan Key Laboratory for Super-microstructure and Ultrafast Process, School of Physics and Electronics, Central South University, Changsha 410083, China}

\author{Qing-Feng Sun}
\affiliation{International Center for Quantum Materials, School of Physics, Peking University, Beijing 100871, China}
\affiliation{Collaborative Innovation Center of Quantum Matter, Beijing 100871, China}
\affiliation{CAS Center for Excellence in Topological Quantum Computation, University of Chinese Academy of Sciences, Beijing 100190, China}
\date{\today}

\begin{abstract}
As a secondary structure of DNA, DNA tetrahedra exhibit intriguing charge transport phenomena and provide a promising platform for wide applications like biosensors, as shown in recent electrochemical experiments. Here, we study charge transport in a multi-terminal DNA tetrahedron, finding that its charge transport properties strongly depend upon the interplay among contact position, on-site energy disorder, and base-pair mismatch. Our results indicate that the charge transport efficiency is nearly independent of contact position in the weak disorder regime, and is dramatically declined by the occurrence of a single base-pair mismatch between the source and the drain, in accordance with experimental results [J. Am. Chem. Soc. {\bf 134}, 13148 (2012); Chem. Sci. {\bf 9}, 979 (2018)]. By contrast, the charge transport efficiency could be enhanced monotonically by shifting the source toward the drain in the strong disorder regime, and be increased when the base-pair mismatch takes place exactly at the contact position. In particular, when the source moves successively from the top vertex to the drain, the charge transport through the tetrahedral DNA device can be separated into three regimes, ranging from disorder-induced linear decrement of charge transport to disorder-insensitive charge transport, and to disorder-enhanced charge transport. Finally, we predict that the DNA tetrahedron functions as a more efficient spin filter compared to double-stranded DNA and opposite spin polarization could be observed at different drains, which may be used to separate spin-unpolarized electrons into spin-up ones and spin-down ones. These results could be readily checked by electrochemical measurements and may help for designing novel DNA tetrahedron-based molecular nanodevices.
\end{abstract}
\maketitle

\section{\label{sec1}Introduction}

Since the original proposal by Eley and Spivey that $\pi$ stacking along the helix axis could provide a natural pathway for conducting electrons in DNA molecules \cite{ddf}, their charge transport properties have been attracting extensive attention among the physics, chemistry, and biology communities \cite{rgen,tch,jcg,fev}, finding that DNA functions as an important candidate for molecular electronics. Recent charge transport experiments have demonstrated a number of fascinating phenomena in double-stranded DNA (dsDNA) by means of advanced experimental techniques \cite{bghl,cbr,lxia,clg,yql,lxi,rzh}. For example, G\"{o}hler {\it et al.} have reported that spin-unpolarized electrons become highly spin polarized when transmitting through a self-assembled monolayer of dsDNA \cite{bghl}. Guo {\it et al.} have proposed a DNA-based molecular rectifier with high rectification ratio upon intercalation of coralyne into a single dsDNA \cite{clg}. Beyond current experimental observations, many interesting phenomena have also been predicted in dsDNA devices \cite{avma,ssm,cts,bta,amg2,hzt,amg3,aagg}, such as the emergence of Majorana zero modes and topological charge pumping. However, when the $\pi$ stacking is perturbed by either base-pair mismatch or oxidative damage, the charge transport through DNA could be significantly decreased, as demonstrated by experimental and theoretical works \cite{so,bgi,pkb,fsh,jhi,uos,ned,mhl,smi}.

\begin{figure}
\includegraphics[width=8.6cm]{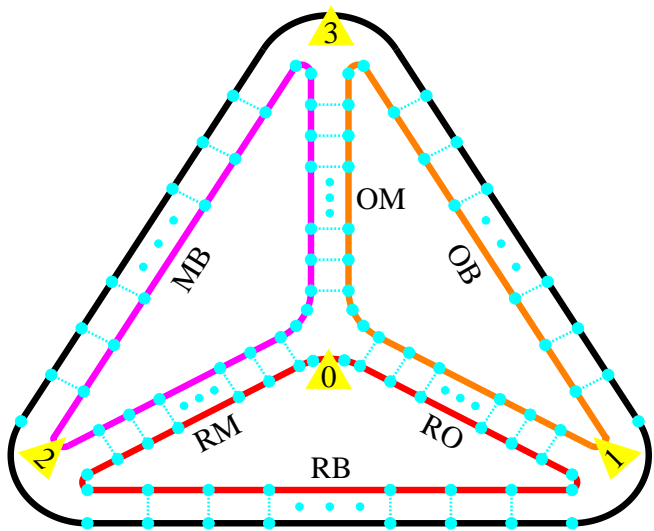}
\caption{\label{fig1} Top view of a DNA tetrahedron composed of six Watson-Crick-paired edges, where two neighboring edges are linked by unpaired hinge nucleobases. This DNA tetrahedron can be self-assembled from four specifically designed single-stranded DNA which are represented by the red, orange, magenta, and black lines. Each double-stranded edge can then be labeled by combining two different lines, including the RO, RM, RB, OM, OB, and MB edges. This three-dimensional DNA structure is contacted by four nonmagnetic electrodes $e_0$, $e_1$, $e_2$, and $e_3$ (see the yellow triangles indicated by arabic numerals 0, 1, 2, and 3). The electrode $e_0$ can move along the RO edge and act as the source, and the other electrodes are fixed at the three bottom vertices ${\rm D}_{j(j=1,2,3)}$ and serve as the drain, just as electrochemical experiments \cite{nluh,clix}. Here, $e_0$ is contacted at the top vertex ${\rm D_0}$ of the DNA tetrahedron with $P=0$ (see text), the big cyan spheres denote nucleobases, and the dotted lines stand for hydrogen bonding within a single base pair. For clarity, the helical structure is not presented.}
\end{figure}

Besides the traditional double helix, DNA can form diverse two- and three-dimensional (3D) nanostructures like DNA tetrahedra \cite{ncse,hyan,rpg1,rpg2,pwkr,smdo,esan}. The DNA tetrahedron consists of six Watson-Crick-paired edges linked by unpaired hinge nucleobases, where each double-stranded edge is named as the combination of two different lines, including the RO, RM, RB, OM, OB, and MB edges, as illustrated in Fig.~\ref{fig1}, which possesses several superior advantages. (i) The DNA tetrahedron can be self-assembled from four well-designed single-stranded DNA (ssDNA) (see the red, orange, magenta, and black lines in Fig.~\ref{fig1}), which could be accomplished in only a few seconds and much simpler compared to the synthesis of other DNA nanostructures, as firstly reported by Turberfield {\it et al.} \cite{rpg1,rpg2}.  (ii) The DNA tetrahedron has high mechanical rigidity and could stay normal to the surface, thus avoiding the crowding effect and sample collision which occur in dense dsDNA monolayers \cite{wz,slit}. (iii) The DNA tetrahedron is capable of entering cells efficiently and can thus deliver cargoes, such as drugs, across cells using its hollow structure \cite{asw,zxia,qlid,hdi}.

Specifically, Fan {\it et al.} have designed a four-terminal tetrahedral DNA device \cite{hpei,nluh,aabi}, where the top vertex ${\rm D}_0$ is connected to a redox molecule (electrode $e_0$) and the three bottom ones ${\rm D}_{j(j=1,2,3)}$ to the surface to immobilize the DNA tetrahedron via thiol groups, as shown in Fig.~\ref{fig1}. By performing electrochemical experiments, the charge transmission in the DNA tetrahedron has been intensively studied and presents distinct characteristics \cite{hpei,nluh,aabi,clix}. (i) When the redox molecule of methylene blue is separated from the surface by either four or thirteen nucleobases, the charge transport efficiency keeps almost the same, implying that the charge transport along the DNA tetrahedron may be insensitive to the position of redox molecule. (ii) When a single base-pair mismatch takes place between the redox molecule and the surface, the charge transport efficiency could diminish dramatically. In addition, many experiments have demonstrated that the DNA tetrahedron provides a promising platform for realizing biosensors which are superior to dsDNA \cite{hpei,nluh,aabi,clix,kjca,ywen,zgem,mli,pso1,pso2}, where the charge transport plays an important role. However, the underlying physics remains unclear regarding charge transport along the DNA tetrahedron.

In this paper, we study theoretically the charge transmission through a multi-terminal DNA tetrahedron as in electrochemical experiments by considering contact position, on-site energy disorder, and base-pair mismatch. Here, the variation of contact position is achieved by moving the source $e_0$ along the RO edge, and the three drains $e_1$, $e_2$, and $e_3$ are fixed at the bottom vertices ${\rm D}_1$, ${\rm D}_2$, and  ${\rm D}_3$, respectively, as shown by the yellow triangles in Fig.~\ref{fig1}. Our results indicate that the charge transport properties of the DNA tetrahedron strongly depend on the competition among contact position, on-site energy disorder, and base-pair mismatch, because of multiple transport pathways. When $e_0$ is contacted at the top vertex ${\rm D}_0$, the transmission spectra at distinct drains are similar but the magnitude of conductance is different. We then focus on the charge transport detected at $e_1$, showing that the charge transmission in the DNA tetrahedron presents several intriguing phenomena when $e_0$ moves along the RO edge. First, the charge transport ability is approximately independent of $e_0$'s position in the weak disorder regime when $e_0$ is separated from $e_1$ by at least a few nucleobases, and is dramatically reduced by the occurrence of a single base-pair mismatch between $e_0$ and $e_1$, consistent with experimental results \cite{nluh,clix}. Second, the charge transport ability could be enhanced monotonically by shifting $e_0$ toward $e_1$ in the strong disorder regime, and be increased when the base-pair mismatch takes place exactly at $e_0$. Third, when $e_0$ moves continuously from ${\rm D}_0$ to $e_1$, the charge transport through the tetrahedral DNA device can be divided into three regimes, ranging from disorder-induced linear decrement of charge transport to disorder-insensitive charge transport, and to disorder-enhanced charge transport which is contrary to the common viewpoint that the transmission ability should become poorer for stronger disorder. Finally, we predict that the DNA tetrahedron could exhibit significant spin-filtering effect, with spin filtration efficiency much larger than dsDNA. In particular, the spin-polarized direction could be opposite at different drains, which may be used to separate spin-unpolarized electrons into spin-up ones and spin-down ones. The underlying physics of all these transport phenomena is analyzed.

The rest of the paper is constructed as follows. Section~\ref{sec2} presents the model Hamiltonian of a multi-terminal DNA tetrahedron and the Green's function. Section~\ref{sec3} shows the numerical results and discussion. Section~\ref{sec3a} studies the effects of contact position and Anderson disorder on charge transport along the DNA tetrahedron, Sec.~\ref{sec3b} considers the base-pair mismatch effect, and Sec.~\ref{sec3c} investigates the spin-filtering effect. Finally, the results are concluded in Sec.~\ref{sec4}.

\section{\label{sec2}MODEL AND METHOD}

The charge transport through a multi-terminal DNA tetrahedron can be described by the model Hamiltonian $\mathcal{H}=\mathcal{H}_ {\rm m}+ \mathcal{H} _{\rm d} +\mathcal{H} _{\rm e}$. The first term, $\mathcal{H}_ {\rm m}$, is the Hamiltonian of an isolated DNA tetrahedron including both double-stranded edges and unpaired hinge nucleobases, which reads \cite{amg1}
\begin{equation}
\begin{aligned}
\mathcal{H}_{\rm m} =& \sum_{j=0}^3 \left\{\sum_{n=1} ^{N_j} [\varepsilon_{jn} c_{jn}^{\dagger} c_{jn}+ t_{n,n+1} ^{(j)} c_{jn}^{\dagger} c_{jn+1}]\right.\\ &+\left. \sum_{s=1} ^{3}\sum_ {n=j_s} ^{j_s+N-2}i t_{\rm so} c_{jn}^ {\dagger} [\sigma_{n}^ {(\eta)}+ \sigma_{n+1} ^{(\eta)}] c_{jn+1}+\mathrm{H.c.}\right\} \\&+ \sum_{\langle jn,j'n' \rangle} \lambda c_{jn}^{\dagger} c_{j'n'}. \label{eq1}
\end{aligned}
\end{equation}
Here, $c_{jn}^{\dagger} =(c_{jn\uparrow} ^{\dagger}, c_{jn \downarrow} ^{\dagger})$ is the creation operator of an electron at site $\{j,n\}$, with $c_{jN_j+1} ^{\dagger}=c_{j1} ^{\dagger}$, $j$ labeling the ssDNA whose length is $N_j$ (see the red, orange, magenta, and black lines in Fig.~\ref{fig1}) and $n$ the nucleobase index (see the big cyan spheres in Fig.~\ref{fig1}). $\varepsilon_{jn}$ is the on-site energy, $t_{n,n+1}^{(j)}$ ($\lambda$) is the intrastrand (interstrand) coupling, and $t_{\rm so}$ the spin-orbit coupling (SOC) parameter. Each ssDNA contains three segments $s=1$, 2, and 3, all of which pair with the complementary ssDNA segment and lead to six self-assembled double-stranded edges (Fig.~\ref{fig1}). $j_s$ is the nucleobase index at which the base pairing begins in the $s$th segment of the $j$th ssDNA and $N$ is the length of the double-stranded edges. The SOC term is expressed as $\sigma_{n} ^{(\eta)}= \sigma_{z} \cos\theta-(-1)^{\eta} \{\sigma_{x} \sin [(n-j_s) \Delta \varphi]-\sigma_{y} \cos [(n-j_s) \Delta \varphi]\} \sin\theta$ \cite{amg1}, where $\sigma_{x, y, z}$ are the Pauli matrices, $\theta$ the space angle between the helical strand and the plane perpendicular to the corresponding helix axis, $\Delta \varphi$ the twist angle between two neighboring base pairs, and $\eta=1,2$ is the strand index of double-stranded edges. $\langle...\rangle$ represents the nearest-neighbor nucleobases between two complementary segments of the $j$th and $j'$th ssDNA.

The second term, $\mathcal{H} _{\rm d}$, describes the dephasing processes during the charge transport in the DNA tetrahedron, which are caused by inelastic scatterings, such as the electron-phonon interaction and the electron-electron interaction. These dephasing processes can be simulated by connecting each nucleobase to a B\"{u}ttiker's virtual electrode \cite{amg1}.

The last term, $\mathcal{H}_{\rm e}$, represents the four nonmagnetic electrodes and their couplings to the DNA tetrahedron (see the yellow triangles in Fig.~\ref{fig1}). The drain electrodes $e_1$, $e_2$, and $e_3$, aiming to anchoring the DNA tetrahedron on the substrate, are contacted at the three bottom vertices ${\rm D}_1$, ${\rm D}_2$, and ${\rm D}_3$, respectively. Their Hamiltonian is expressed as:
\begin{equation}
\mathcal{H} _{\rm e}^{(j)} =\sum_ {k} [\varepsilon_{jk} a_{j k}^ {\dagger} a_{j k}+ \tau a_{j k} ^{\dagger}  (c_{j1} +c_{j N_j})+ \mathrm{H.c.}]. \label{eq2}
\end{equation}
Here, $a_{j k}^{\dagger}=(a_{j k\uparrow} ^{\dagger}, a_{j k\downarrow} ^{\dagger})$ is the creation operator of mode $k$ in the drain electrode $e_{j}$, and $\tau$ is the coupling between the DNA tetrahedron and the electrodes, with $j=1$, 2, and 3. Notice that each electrode $e_j$ is connected to both ends of the $j$th ssDNA. For the source electrode $e_0$, it can move along the RO edge and the Hamiltonian is:
\begin{equation}
\mathcal{H} _{\rm e}^{(0)} =\sum_ {k} [\varepsilon_{0k} a_{0 k}^ {\dagger} a_{0 k}+ \tau a_{0 k} ^{\dagger}  (c_{0P} +c_{0 P+1})+ \mathrm{H.c.}], \label{eq3}
\end{equation}
where $e_0$ is connected to the $P$th and $P+1$th nucleobases of the $0$th ssDNA, with $P$ being the contact position and $c_{00} ^\dagger \equiv c_{0N_0} ^\dagger$. Figure~\ref{fig1} presents a tetrahedral DNA device where $e_0$ is contacted at ${\rm D}_0$ with $P=0$. We emphasize that when an electrode is intercalated between two neighboring nucleoases, the direct connection between these two nucleobases is replaced by indirect connection via this electrode and the corresponding intrastrand coupling disappears.

By employing the Landauer-B\"{u}ttiker formula together with the nonequilibrium Green's function, the current flowing through the $p$th electrode (real or virtual) is expressed as \cite{sd}:
\begin{equation}
I_{p}=\frac{2e^{2}}{h} \sum_{q} T_{p, q}\left(V_{q}-V_{p}\right), \label{eq4}
\end{equation}
where $V_q$ and $V_p$ are, respectively, the voltages applied in the $q$th and $p$th electrodes, and
\begin{equation}
T_{p,q}=\operatorname{Tr}[\mathbf{\Gamma}_{p} \mathbf{G}^{r} \mathbf{\Gamma}_{q} \mathbf{G}^{a}] \label{eq5}
\end{equation}
is the transmission coefficient from the $q$th electrode to the $p$th one. The Green's function $\mathbf{G} ^{r} (E)= [\mathbf{G} ^{a} (E)]^ {\dagger} =[E \mathbf{I} -\mathbf{H}_ {\rm m}- \sum_{p} \mathbf{\Sigma} _{p} ^{r}]^ {-1}$ and the linewidth function $\mathbf{\Gamma} _{p}= i [\mathbf{\Sigma} _{p} ^{r}- (\mathbf{\Sigma} _{p} ^{r}) ^{\dagger}]$, with $E$ the electron energy, $\mathbf{H}_ {\rm m}$ the Hamiltonian of the isolated DNA tetrahedron in the site representation, and $\mathbf{\Sigma}_{p}^{r}$ the retarded self-energy owing to the coupling to the $p$th electrode. In the wide-band limit, the retarded self-energy is taken as $\Sigma_{p}^{r}=-i \Gamma / 2$ for the real electrodes and $\Sigma_{p}^{r}=-i \Gamma_{d} / 2$ for the virtual ones, with $\Gamma$ being the coupling strength between the real electrodes and the DNA tetrahedron and $\Gamma_{d}$ the dephasing parameter. Here, we consider a small bias voltage $V_b$ between the source and the drain, with $V_{e_0}=V_b$ for $e_0$ and $V_{e_j}=0$ for $e_{j (j=1,2,3)}$, and the net current flowing through each virtual electrode is zero. Under these boundary conditions, the voltages of the virtual  electrodes can be calculated from Eq.~(\ref{eq4}). Then, the conductance of all the drains $e_{j(j=1, 2, 3)}$ can be obtained as:
\begin{equation}
G_{j}=\frac{2e^{2}}{h} \sum_{q} T_{e_j, q} \frac{V_q} {V_b}.
\label{eq6}
\end{equation}

\section{\label{sec3} RESULTS AND DISCUSSION}

We consider the case that all the double-stranded edges are identical homogeneous dsDNA molecules, e.g., poly(G)-poly(C), with G the guanine and C the cytosine. The on-site energy and the intrastrand coupling are set to $\varepsilon_{jn} = \varepsilon_{\rm G}$ and $t_ {n,n+1} ^{(j)} =t_{\rm G}$ for the first strand, $\varepsilon_{jn} = \varepsilon_{\rm C}$ and $t_{n, n+1} ^{(j)} =t_{\rm C}$ for the second strand, and $\varepsilon_{jn} = \varepsilon_{\rm G}$ for the unpaired hinge nucleobases. As the paired nucleobase could be different from the unpaired one in the same ssDNA, the intrastrand coupling between two neighboring G and C nucleobases is taken as $t_ {n,n+1} ^{(j)} =t_{\rm GC}$. According to first-principles calculations \cite{hzx,kse,lgd}, these model parameters are set to $\varepsilon_{\rm G}=0$ (energy reference point), $\varepsilon_{\rm C} =0.34$, $t_{\rm G}= t_{\rm GC}=0.08$, $t_{\rm C}=-0.12$, and $\lambda= -0.17$, with the unit in eV. The SOC strength is set to $t_{\rm so} = 0.01$, and the structural parameters to $\theta \approx0.66$ and $\Delta\phi = \pi/5$ \cite{amg1}. The size of the DNA tetrahedron is set to $N=17$, $N_0=57$, and $N_{1} =N_{2} =N_{3} =55$, just the same as the experiment \cite{nluh}. The coupling parameter is $\Gamma=1$ for the real electrodes, and the dephasing strength is $\Gamma_{d} =10^{-4}$ for the virtual ones because the DNA tetrahedron presents higher mechanical rigidity compared to traditional dsDNA molecules. Since the hinge nucleobases around the four vertices ${\rm D}_{j(j=0,1,2,3)}$ are unpaired and cannot form a well-defined secondary structure, the corresponding intrastrand coupling is estimated to half of its initial value and the SOC is neglected. As a result, the SOC and the interstrand coupling only exist in the double-stranded edges.

\subsection{\label{sec3a} Contact position and disorder effects on charge
transport along tetrahedral DNA devices}

\begin{figure}
\includegraphics[width=8.6cm]{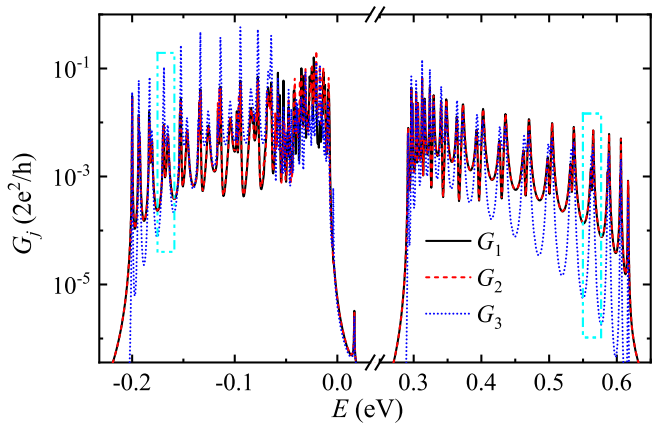}
\caption{\label{fig2} Charge transport along the tetrahedral DNA device in the absence of on-site energy disorder when the source electrode $e_0$ is contacted at its top vertex with $P=0$. Energy-dependent conductance $G_1$ detected at the drain electrode $e_1$ (black-solid line), $G_2$ at $e_2$ (red-dashed line), and $G_3$ at $e_3$ (blue-dotted line).}
\end{figure}

We first study the charge transport properties of the tetrahedral DNA device in the absence of on-site energy disorder when the source electrode is contacted at its top vertex ${\rm D}_0$ with $P=0$ (Fig.~\ref{fig1}). In this situation, all the real electrodes are connected to the four nearly equivalent vertices ${\rm D}_{j(j=0,1,2,3)}$ of the DNA tetrahedron and leads to similar transmission spectra probed at different drain electrodes, as can be seen from Fig.~\ref{fig2}, where the conductances $G_1$, $G_2$, and $G_3$ of the three drain electrodes are plotted as a function of the electron energy $E$. By inspecting Fig.~\ref{fig2}, it is clear that all the transmission spectra always consist of two electronic bands separated by an energy gap. Besides, a number of pronounced transmission peaks and valleys can be found in each band and the density of peaks/valleys becomes larger when $E$ is close to the energy gap, owing to the SOC effect. These features are similar to a single dsDNA molecule \cite{amg1,sro}, because the DNA tetrahedron consists of six identical dsDNA molecules.

It is interesting that for both bands of the DNA tetrahedron, each original transmission peak is usually split into a pair of minor peaks separated by a dip (see the two cyan rectangles in Fig.~\ref{fig2}) and consequently there are $2(N-1)$ transmission peaks in the right band, which is different from a single dsDNA \cite{amg1,sro} and can be understood from multiple transport pathways between the source and the drain. Let us take electron flowing from $e_0$ to $e_1$ as an example. In this case, the electrons can transport via multiple pathways, including
\begin{equation}
\begin{aligned}
&{\rm D_0} \rightarrow {\rm RO} \rightarrow {\rm D_1}, \\
&{\rm D_0} \rightarrow {\rm RM} \rightarrow {\rm D_2} \rightarrow  {\rm RB} \rightarrow {\rm D_1}, \\
&{\rm D_0} \rightarrow {\rm OM} \rightarrow {\rm D_3} \rightarrow  {\rm OB} \rightarrow {\rm D_1}, \\
&{\rm D_0} \rightarrow {\rm RM} \rightarrow {\rm D_2} \rightarrow {\rm MB} \rightarrow {\rm D_3} \rightarrow  {\rm OB} \rightarrow {\rm D_1}, \\
\text{and} \\
&{\rm D_0} \rightarrow {\rm OM} \rightarrow {\rm D_3} \rightarrow {\rm MB} \rightarrow {\rm D_2} \rightarrow  {\rm RB} \rightarrow {\rm D_1}. \label{eq7}
\end{aligned}
\end{equation}
Notice that the second pathway is equivalent to the third one, because these two pathways contain the same number of the unpaired nucleobases and all the double-stranded edges are identical. In the strong scattering regime where $E$ is far away from $\varepsilon_{\rm G}$, the unpaired nucleobases function as strong potential barriers/wells and the electrons will mainly propagate along relatively short pathways, i.e., the former three pathways, giving rise to the splitting of an original transmission peak into a pair of minor peaks. When $E$ is shifted toward the energy gap, the scattering from the unpaired nucleobases is gradually weakened, and correspondingly the electron transport through the second and third pathways becomes more and more pronounced. As a result, the quantum interference among different pathways should be progressively enhanced, leading to the increment of the left (right) minor peak in the right (left) band and further separation of these two minor peaks (see the right band and the other energy region distant from the energy gap in the left band in Fig.~\ref{fig2}). While in the weak scattering regime where $E$ locates in the vicinity of $\varepsilon_{\rm G}$ with $|E- \varepsilon_{\rm G}| <|t_2|$, the scattering from the unpaired nucleobases is weak and the electrons can propagate through more pathways, including the latter two longer pathways in Eq.~(\ref{eq7}). Consequently, an original transmission peak will be divided into several minor peaks and the charge transport along the DNA tetrahedron becomes complicated when $E$ is close to the on-site energy of the unpaired nucleobases.

Despite the similarity of the transmission spectra at different drain electrodes, the transmission profiles depend on the drain position, which is related to the source position. One can see from Fig.~\ref{fig1} that the electrons can transport from $e_0$ to $e_3$ mediated by multiple pathways as well, such as ${\rm D_0} \rightarrow {\rm OM} \rightarrow {\rm D_3}$. Notice that this shortest pathway differs from the one, ${\rm D_0} \rightarrow {\rm RO} \rightarrow {\rm D_1}$, regarding electron flowing from $e_0$ to $e_1$, because the number of unpaired nucleobases in these two pathways is different, leading to distinct magnitude of $G_1$ and $G_3$ (see the black-solid and blue-dotted lines in Fig.~\ref{fig2}). By contrast, the former three pathways in Eq.~(\ref{eq7}) are equivalent to those of electron flowing from $e_0$ to $e_2$ and thus there is no observable difference between $G_1$ and $G_2$ when $E$ is far away from $\varepsilon_{\rm G}$ (see the black-solid and red-dashed lines in Fig.~\ref{fig2}). The relationship among these transmission spectra can also be understood by analyzing the symmetry of the tetrahedral DNA device. By inspecting Fig.~\ref{fig1}, it is clear that in the presence of the source electrode, the mirror symmetry is broken with respect to the normal plane across the RO (RM) edge, whereas it is almost preserved with respect to the normal plane through the OM edge. Therefore, the curve of $G_3-E$ is different from both $G_1-E$ and $G_2-E$, and the latter two curves almost overlap.

\begin{figure}
\includegraphics[width=8.6cm]{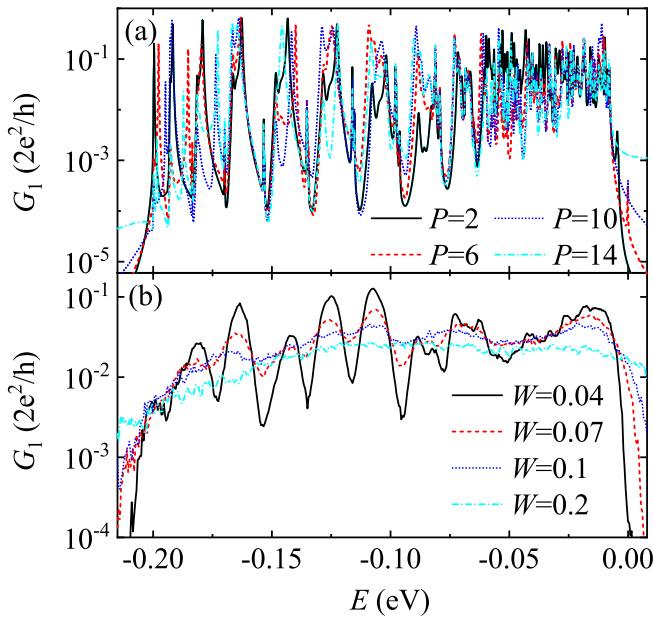}
\caption{\label{fig3} Charge transport along the tetrahedral DNA device by considering contact position and on-site energy disorder. (a) $G_1$ versus $E$ for typical contact positions $P$ in the absence of on-site energy disorder, $W=0$. (b) $G_1$ versus $E$ for different disorder strengths $W$ with $P=6$.}
\end{figure}

\begin{figure}
\includegraphics[width=8.6cm]{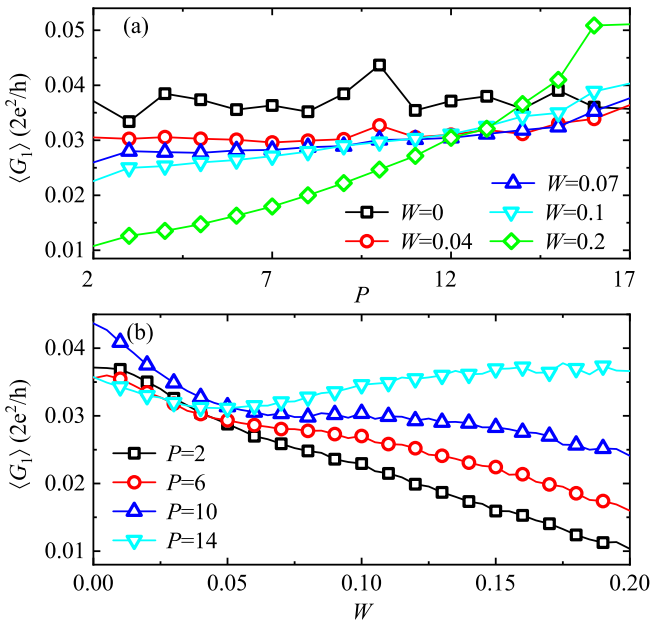}
\caption{\label{fig4} Charge transport along the tetrahedral DNA device by considering contact position and on-site energy disorder. (a) Averaged conductance $\langle G_1\rangle$ versus contact position $P$ for different disorder strengths $W$. (b) $\langle G_1\rangle$ versus $W$ for typical values of $P$.}
\end{figure}

We then consider the effect of contact position and on-site energy disorder by exploring the charge transport at the drain electrode $e_1$, and focus on  the left electronic band for clarity. Figure~\ref{fig3}(a) shows $G_1$ versus $E$ for typical contact positions $P$, which is achieved by moving the source electrode $e_0$ along the RO edge, in the absence of on-site energy disorder. In this case, the former three pathways in Eq.~(\ref{eq7}) are changed into
\begin{equation}
\begin{aligned}
&P \rightarrow {\rm \overline{RO}} \rightarrow {\rm D_1}, \\
&P \rightarrow {\rm \overline{RO}} \rightarrow {\rm D_0} \rightarrow {\rm RM} \rightarrow {\rm D_2} \rightarrow  {\rm RB} \rightarrow {\rm D_1}, \\
\text{and} \\
&P \rightarrow {\rm \overline{RO}} \rightarrow {\rm D_0} \rightarrow {\rm OM} \rightarrow {\rm D_3} \rightarrow  {\rm OB} \rightarrow {\rm D_1}, \label{eq8}
\end{aligned}
\end{equation}
where ${\rm \overline{RO}}$ denotes part of the RO edge. It is clear that the shortest pathway, $P \rightarrow {\rm \overline{RO}} \rightarrow {\rm D}_1$, does not contain any unpaired nucleobase and the resulting scattering vanishes when the electrons pass through this pathway, which could considerably enhance the transmission ability at $e_1$ [see the black-solid line in Fig.~\ref{fig2} and all the lines in Fig.~\ref{fig3}(a)]. In particular, although this shortest pathway becomes shorter when $e_0$ moves toward $e_1$, the transmission spectra are quite similar for different $P$'s and the number of transmission peaks is almost independent of $P$ [Fig.~\ref{fig3}(a)], another signature of multiple transport pathways for electron flowing from $e_0$ to $e_1$. Further studies demonstrate that the averaged conductance $\langle G_1 \rangle$, obtained from the left band, is approximately independent of $P$ [see the black squares in Fig.~\ref{fig4}(a)], in accordance with the experiment \cite{nluh}. When $e_0$ deviates from the top vertex, the second and third pathways in Eq.~(\ref{eq8}) possess different number of unpaired nucloebases and correspondingly these two pathways are not equivalent. As a result, each original transmission peak will be split into more than two minor peaks [Fig.~\ref{fig3}(a)], and the charge transport along the DNA tetrahedron becomes more complicated compared to the case of $P=0$.

Notice that the DNA tetrahedron can also be composed of inhomogeneous dsDNA molecules. And the electrochemical experiments were carried out in phosphate buffer solution \cite{hpei,nluh,aabi,clix}, where counterions and water molecules may adsorb randomly around the DNA tetrahedron. To simulate inhomogeneous dsDNA molecules and the environmental effect, Anderson disorder is considered by adding a random variable $w_{jn}$ in the on-site energy $\varepsilon_{jn}$ \cite{yuz}, with $w_{jn}$ uniformly distributed within the range $[-\frac W 2,\frac W 2]$ and $W$ the disorder degree. Figure~\ref{fig3}(b) shows $G_1$ versus $E$ for typical disorder degrees $W$ with $P=6$, which is calculated from an ensemble of 2000 disorder configurations. Once Anderson disorder is introduced, the electrons will experience stronger scattering when transmitting along longer pathways and preferentially propagate through the shortest pathway, leading to dramatic decrement in the number of transmission peaks [see the black-solid line in Fig.~\ref{fig3}(b)]. The larger Anderson disorder is, the stronger scattering the electrons suffer. As a result, both magnitude and oscillation amplitude of $G_1$ decrease with increasing $W$. Nevertheless, in relatively weak disorder regime, $G_1$ can increase with $W$ around the transmission valleys [see the black-solid, red-dashed, and blue-dotted lines in Fig.~\ref{fig3}(b)], because of the increment of electronic states.

To further demonstrate the interplay between the contact position and the on-site energy disorder, the averaged conductance is calculated
\begin{equation}
\langle G_1\rangle =\frac{1} {\Omega} \int_{\Omega} G_1 dE, \label{eq9}
\end{equation}
where $\Omega$ denotes the left electronic band. Figure~\ref{fig4}(a) displays $\langle G_1\rangle$ versus $P$ for several disorder degrees $W$. It is clear that the dependence of $\langle G_1\rangle$ on $P$ is determined by both $W$ and $P$. (i) In the absence of Anderson disorder, $\langle G_1\rangle$ fluctuates around a certain value and is approximately independent of $P$ [see the black squares in Fig.~\ref{fig4}(a)], because of the multiple transport pathways. (ii) In the weak disorder regime, both the second and third pathways in Eq.~(\ref{eq8}) become longer when $e_0$ moves toward $e_1$ and the scattering from these two pathways is gradually enhanced. When $e_0$ is distant from $e_1$, the electrons can propagate through the three pathways in Eq.~(\ref{eq8}), and thus $\langle G_1\rangle$ is insensitive to $P$ within the range $P \in[2, 14]$ [see the red circles and blue-up triangles in Fig.~\ref{fig4}(a)], in good agreement with the experiment \cite{nluh}. By contrast, when $e_0$ is close to $e_1$, the electron transport is mainly mediated by the shortest pathway which becomes shorter by shifting $e_0$ toward $e_1$, and thus the scattering from this pathway gradually weakens and $\langle G_1\rangle$ increases almost linearly with $P$. (iii) In the strong disorder regime, the scattering from longer pathways is so strong that the electron transport through these pathways is negligible and is then mainly mediated by the shortest pathway. By shifting $e_0$ toward $e_1$, the shortest pathway becomes shorter and the corresponding scattering progressively decays, leading to monotonic increasing of $\langle G_1\rangle$ with $P$ in the whole range $P \in[1, 16]$ [see the cyan-down triangles and green diamonds in Fig.~\ref{fig4}(a)].

In addition, one can see from Fig.~\ref{fig4}(a) that in the presence of Anderson disorder, different curves of $\langle G_1\rangle -P$ intersect at about $P_c=12$, where $\langle G_1\rangle $ decreases with $W$ for $P<P_c$ and contrarily increases with $W$ for $P>P_c$. To further demonstrate the counterintuitive phenomenon of the enhancement of the transmission ability by Anderson disorder, Fig.~\ref{fig4}(b) plots $\langle G_1\rangle $ versus $W$ for typical contact positions $P$. One can see that the dependence of $\langle G_1\rangle $ on $W$ is also determined by both $P$ and $W$. When $P=2$, $\langle G_1\rangle $ decreases almost linearly with increasing $W$ [see the black squares in Fig.~\ref{fig4}(b)]. This is different from a single dsDNA molecule where the conductance decays exponentially with the disorder degree \cite{gam}, stemming from the fact that the three pathways in Eq.~(\ref{eq8}) are approximately equivalent as $e_0$ is apart from the top vertex by only two nucleobases.

When $e_0$ is distant from the top vertex, however, the curves of $\langle G_1 \rangle- W$ are different and could be divided into three parts in general. In the weak disorder regime, $\langle G_1\rangle $ decreases quickly with $W$ as expected [see the beginning parts of the red circles, blue-up triangles, and cyan-down triangles in Fig.~\ref{fig4}(b)], because the electron transport through all the pathways is declined caused by Anderson localization. When $W$ is increased and reaches the intermediate regime, the scattering becomes stronger, but $\langle G_1\rangle $ decreases slowly with $W$ or is immune to $W$ [see the middle parts of the red circles and blue-up triangles in Fig.~\ref{fig4}(b)]. In particular, $\langle G_1\rangle$ can even increase with $W$ when $e_0$ is close to $e_1$ [see the middle part of the cyan-down triangles in Fig.~\ref{fig4}(b)], which has been reported in other one- and two-dimensional systems \cite{gam1,eli} but the physical mechanism is different. Notice that with increasing $W$, the increasing rate of scattering from longer pathways is faster and the electrons will preferentially propagate through the shortest pathway. In other words, the probability of electron transmission through the shortest pathway increases with $W$, which competes with the enhanced scattering from this pathway. When $e_0$ moves toward $e_1$, the electron transmission probability through the shortest pathway is gradually enhanced, whereas the scattering from this pathway is declined. As a result, in the intermediate disorder regime, the electron flowing from $e_0$ to $e_1$ can be categorized into three mechanisms with increasing $P$, ranging from disorder-induced suppression of charge transport to disorder-insensitive charge transport, and to disorder-enhanced charge transport. In the strong disorder regime, the electron transmission through the shortest pathway is dramatically suppressed and thus $\langle G_1\rangle $ decreases quickly with $W$ again, regardless of the contact position.

\subsection{\label{sec3b} Base-pair mismatch effect on charge transport along tetrahedral DNA devices}

\begin{figure}
\includegraphics[width=7.6cm]{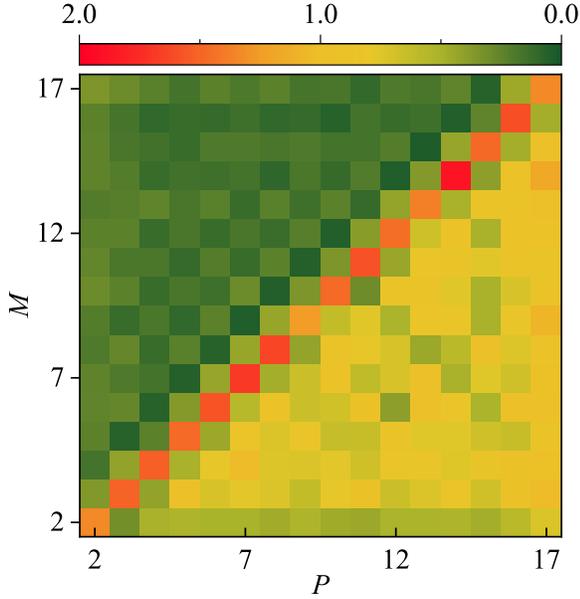}
\caption{\label{fig5} Charge transport along the tetrahedral DNA device by considering contact position and a single base-pair mismatch in the RO edge. A contour plot of the ratio $\langle G_1 \rangle_{\rm MM} / \langle G_1 \rangle_{\rm WM}$ as functions of contact position $P$ and base-pair mismatch site $M$. Here, $\langle G_1\rangle_{\rm MM}$ corresponds to the averaged conductance in the presence of base-pair mismatch and $\langle G_1 \rangle_{\rm WM}$ to the one without any base-pair mismatch.}
\end{figure}

Next, we investigate the interplay between the contact position and a single base-pair mismatch in the RO edge. This base-pair mismatch is introduced by considering a C-thymine (T) mismatch as in the electrochemical experiment \cite{nluh}, which occurs by replacing a G nucleobase with a T one. The on-site energy is changed from $\varepsilon_{0M}= \varepsilon_{\rm G}$ to $\varepsilon_{0M}= \varepsilon_{\rm T}=0.5$ eV correspondingly \cite{hzx,kse,lgd}, with $M$ the mismatched site. Previous experimental and theoretical works have shown that when a base-pair mismatch takes place, both $\pi$ stacking and hydrogen bonding, neighboring to the mismatched site, are dramatically declined, owing to strong fluctuation of unstable mismatched base pair \cite{so,bgi,pkb,fsh,jhi,uos,ned,mhl,smi}. Then, both the intrastrand and interstrand couplings, linking to the T nucleobase, are reduced by one order of magnitude. As a result, the electrons could be considerably reflected by the C-T mismatched base pair.

Figure~\ref{fig5} presents the ratio $\langle G_1 \rangle_{\rm MM} / \langle G_1\rangle _{\rm WM}$ as functions of $P$ and $M$. Here, $\langle G_1 \rangle_{\rm MM}$ denotes the averaged conductance, calculated from the left band, in the presence of a single C-T mismatch and $\langle G_1 \rangle_{\rm WM}$ refers to the one without any base-pair mismatch. It clearly appears that the ratio $\langle G_1 \rangle_{\rm MM} / \langle G_1\rangle _{\rm WM}$ depends strongly on the relative position of the source and the C-T base pair, and can mainly be divided into three regions, i.e., $M>P$, $M=P$, and $M<P$. For $M>P$ where the C-T base pair locates between $e_0$ and $e_1$, the electron transmission through the shortest pathway, $P \rightarrow {\rm \overline{RO}} \rightarrow {\rm D}_1$, will be considerably declined due to strong scattering at the C-T base pair. As a result, the transmission ability at $e_1$ can be dramatically decreased by the C-T base pair, leading to a large green area with small ratio $\langle G_1 \rangle_{\rm MM} / \langle G_1\rangle _{\rm WM} \in (0,0.5)$, as can be seen from the upper left part in Fig.~\ref{fig5}. This is consistent with the experiment \cite{nluh,clix}. For $M<P$ where the C-T base pair locates between $e_0$ and ${\rm D}_0$, the electron transmission through longer pathways, such as the second and third pathways in Eq.~(\ref{eq8}), will be decreased but less affected by the C-T base pair compared to the shortest pathway, because there already exist other potential barriers of the unpaired hinge nucleobases in longer pathways. Consequently, a yellow area with relatively large ratio $\langle G_1 \rangle_{\rm MM} / \langle G_1\rangle _{\rm WM} \in (0.5,1)$ is observed in the lower right part of Fig.~\ref{fig5}. It is interesting that when the source is contacted at the mismatched site with $M=P$, the electrons are dramatically reflected by the T nucleobase and the majority of electrons are injected into the G one, leading to significant increment of electron transmission through the shortest pathway and the C-T mismatch-induced enhancement of conductance at $e_1$.

\subsection{\label{sec3c} Spin-filtering effect of tetrahedral DNA devices}

\begin{figure}
\includegraphics[width=8.6cm]{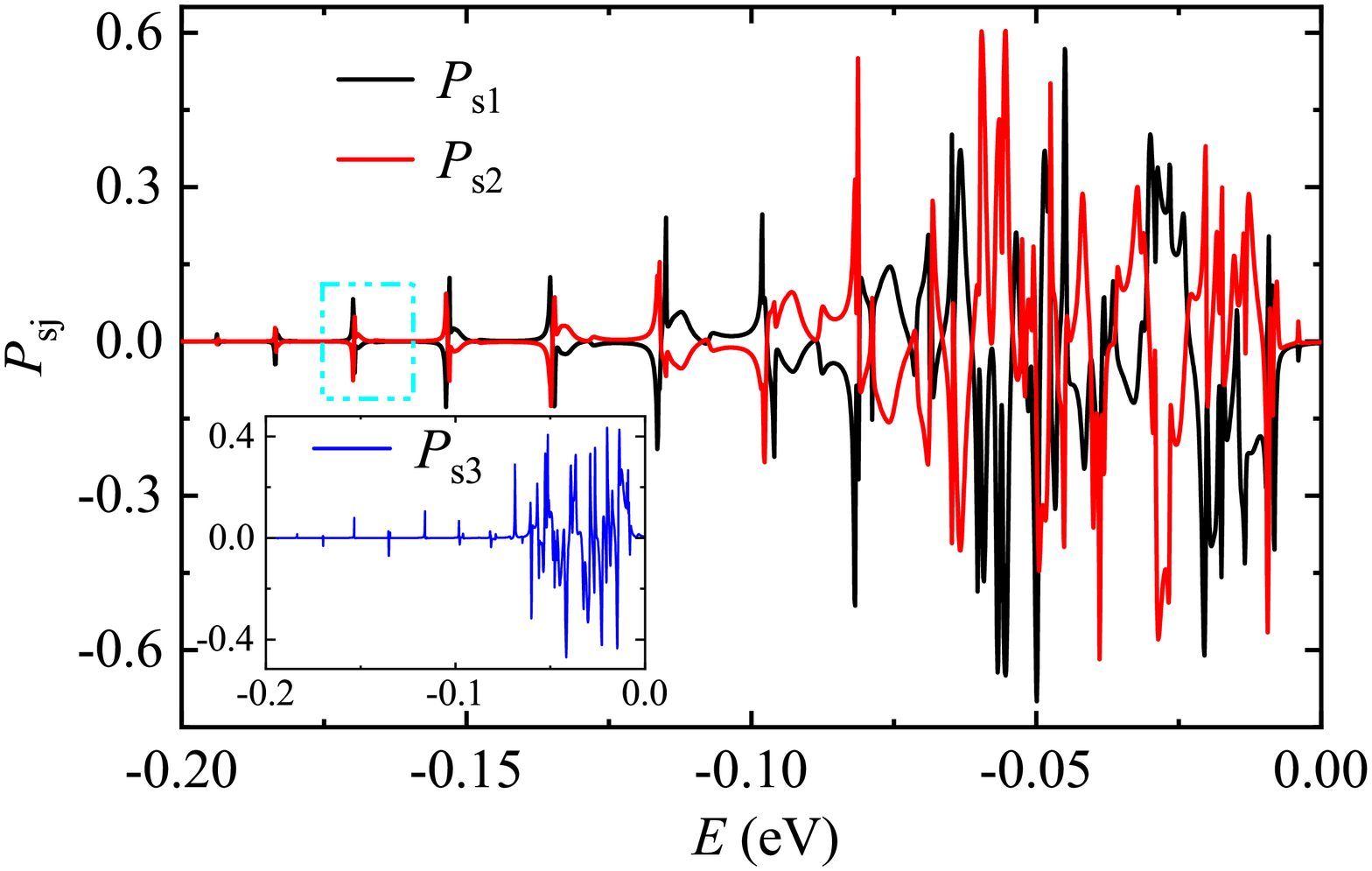}
\caption{\label{fig6} Spin transport along the tetrahedral DNA device when the source electrode $e_0$ is contacted at the top vertex with $P=0$. Energy-dependent spin polarization $P_{{\rm s}1}$ at the drain electrode $e_1$ (black-solid line) and $P_{{\rm s}2}$ at $e_2$ (red-solid line). The inset shows the spin polarization $P_{{\rm s}3}$ at $e_3$.}
\end{figure}

Finally, we study the spin-polarized electron transport along the DNA tetrahedron, where the spin polarization at the $j$th drain is defined as:
\begin{equation}
P_{{\rm s}j}=\frac{G_{j\uparrow} -G_{j\downarrow}} {G_{j\uparrow} +G_{j\downarrow}}. \label{eq10}
\end{equation}

Figure~\ref{fig6} shows $P_{{\rm s}1}$ at $e_1$ (black-solid line) and $P_{{\rm s}2}$ at $e_2$ (red-solid line) as a function of the electron energy $E$, while the inset refers to $P_{{\rm s}3}$ versus $E$ at $e_3$. One can see from Fig.~\ref{fig6} that the spin transport through the DNA tetrahedron has several intriguing phenomena, which are different from dsDNA molecules \cite{bghl,amg1} and DNA hairpins \cite{pjh}. (i) The DNA tetrahedron exhibits pronounced spin filtering effect, regardless of specific drain position. Although each double-stranded edge contains seventeen base pairs with $N=17$, the spin polarization can reach 70.0\% at $e_1$, 61.8\% at $e_2$, and 47.0\% at $e_3$, which is comparable to long dsDNA molecules with length $N=78$ \cite{bghl,amg1}. This large spin polarization further demonstrates the existence of longer transport pathways in the DNA tetrahedron and leads to multiple pathways in this system, because the spin polarization of chiral molecules increases with their molecular length \cite{bghl,amg1}. (ii) The spin polarization oscillates dramatically with increasing $E$, and the closer to the energy gap, the higher the oscillation frequency, which are also independent of the drain position. When the spin polarization reaches the local maximum, the spin-polarized direction could be reversed by increasing $E$ slightly and the spin polarization quickly arrives at the local minimum (see the black-solid line in the cyan rectangle of Fig.~\ref{fig6}), which is accompanied by the splitting of an original transmission peak into two minor peaks (see the black-solid line in the same rectangle of Fig.~\ref{fig2}). This stems from the quantum interference among different pathways and provides an alternative route to reverse the spin-polarized direction by slightly tuning the Fermi energy, instead of flipping the handedness of chiral molecules \cite{amg1}.

Despite the similarity of the spin-polarized profiles at different drain electrodes, the spin polarization considerably depends on the drain position. Although the curves of $G_1-E$ and $G_2-E$ are superimposed upon each other when $E$ is distant from the energy gap (see the black-solid and red-dashed lines in Fig.~\ref{fig2}), the spin-polarized direction at $e_1$ is opposite to that at $e_2$ in almost the whole energy band (see the black- and red-solid lines in Fig.~\ref{fig6}), arising from the nearly preserved mirror symmetry with respect to the normal plane through the OM edge as discussed above. In this sense, the DNA tetrahedron may be used to efficiently separate spin-unpolarized electrons into spin-up electrons and spin-down ones. When spin-up electrons are detected at $e_1$, the spin-down ones could be observed at $e_2$, and vice versa. This theoretical prediction could be readily checked by, e.g., electrochemical experiments \cite{hpei,nluh,aabi,clix}.

\section{\label{sec4} Conclusion}

In summary, the charge transmission in a multi-terminal DNA tetrahedron has been studied by considering contact position, on-site energy disorder, and base-pair mismatch. The variation of contact position is achieved by moving the source along one edge of the DNA tetrahedron, while keeping all the three drains fixed at the bottom vertices. Our results indicate that the charge transport properties of the DNA tetrahedron strongly depend on the competition among contact position, on-site energy disorder, and base-pair mismatch. (i) The dependence of charge transport efficiency on the contact position relies on the disorder. In the weak disorder regime, the charge transport efficiency is approximately independent of the contact position. While in the strong disorder regime, the charge transport efficiency increases monotonically by shifting the source toward the drain. The larger the disorder strength, the faster the increasing rate. (ii) The dependence of charge transport efficiency on the disorder is determined by the contact position. When the source is far from the drain, the charge transport efficiency decreases almost linearly with the disorder strength. When the source is close to the drain, the charge transport efficiency could increase with the disorder strength. (iii) The dependence of charge transport efficiency on the disorder is also determined by the contact position. When the base-pair mismatch occurs between the source and the drain, the charge transport efficiency could be dramatically declined. When the base-pair mismatch occurs at the contact position, the charge transport efficiency could be enhanced. (iv) Finally, we predict that the DNA tetrahedron could behave as a more efficient spin filter compared to double-stranded DNA. In particular, opposite spin polarization could be observed at different drains. This may allow for designing a spin splitter, where spin-up electrons accumulate in one drain and spin-down electrons in another drain.

\section*{Acknowledgments}

This work is supported by the National Natural Science Foundation of China (Grants No. 11874428, No. 11874187, and No. 11921005) and the National Key Research and Development Program of China (Grant No. 2017YFA0303301).


\begin{references}
\bibitem{ddf} D. D. Eley and D. I. Spivey, {Trans. Faraday Soc.} {\bf58}, 411 (1962).
\bibitem{rgen} R. G. Endres, D. L. Cox, and R. R. P. Singh, {Rev. Mod. Phys.} {\bf76}, 195 (2004).
\bibitem{tch} \textsl{Charge Migration in DNA: Perspectives from Physics, Chemistry, and Biology}, edited by T. Chakraborty (Springer, New York, 2007).
\bibitem{jcg} J. C. Genereux and J. K. Barton, {Chem. Rev.} {\bf110}, 1642 (2010).
\bibitem{fev} F. Evers, R. Koryt\'{a}r, S. Tewari, and J. M. van Ruitenbeek, {Rev. Mod. Phys.} {\bf92}, 035001 (2020).


\bibitem{bghl} B. G\"{o}hler, V. Hamelbeck, T. Z. Markus, M. Kettner, G. F. Hanne, Z. Vager, R. Naaman, and H. Zacharias, {Science} {\bf 331}, 894 (2011).
\bibitem{cbr} C. Bruot, J. L. Palma, L. Xiang, V. Mujica, M. A. Ratner, and N. Tao, {Nat. Commun.} {\bf6}, 8032 (2015).
\bibitem{lxia} L. Xiang, J. L. Palma, C. Bruot, V. Mujica, M. A. Ratner, and N. Tao, {Nat. Chem.} {\bf7}, 221 (2015).
\bibitem{clg} C. Guo, K. Wang, E. Zerah-Harush, J. Hamill, B. Wang, Y. Dubi, and B. Xu, {Nat. Chem.} {\bf 8}, 484 (2016).
\bibitem{yql} Y. Li, L. Xiang, J. L. Palma, Y. Asai, and N. Tao, {Nat. Commun.} {\bf 7}, 11294 (2016).
\bibitem{lxi} L. Xiang, J. L. Palma, Y. Li, V. Mujica, M. A. Ratner, and N. Tao, {Nat. Commun.} {\bf 8}, 14471 (2017).
\bibitem{rzh} R. Zhuravel, H. Huang, G. Polycarpou, S. Polydorides, P. Motamarri, L. Katrivas, D. Rotem, J. Sperling, L. A. Zotti, A. B. Kotlyar, J. C. Cuevas, V. Gavini, S. S. Skourtis, and D. Porath, Nat. Nanotechnol. {\bf 15}, 836 (2020).


\bibitem{avma} A. V. Malyshev, {Phys. Rev. Lett.} {\bf98}, 096801 (2007).
\bibitem{ssm} S. S. Mallajosyula and S. K. Pati, {Phys. Rev. Lett.} {\bf98}, 136601 (2007).
\bibitem{cts} C.-T. Shih, S. Roche, and R. A. R\"{o}mer, {Phys. Rev. Lett.} {\bf100}, 018105 (2008).
\bibitem{bta} B. Tan, M. Hodak, W. Lu, and J. Bernholc, {Phys. Rev. B} {\bf 92}, 075429 (2015).
\bibitem{amg2} A.-M. Guo and Q.-F. Sun, {Phys. Rev. B} {\bf 95}, 155411 (2017).
\bibitem{hzt} H.-Z. Tang, Q.-F. Sun, J.-J. Liu, and Y.-T. Zhang, {Phys. Rev. B} {\bf99}, 235427 (2019).
\bibitem{amg3} A.-M. Guo, P.-J. Hu, X.-H. Gao, T.-F. Fang, and Q.-F. Sun, {Phys. Rev. B} {\bf 102}, 155402 (2020).
\bibitem{aagg} A. Aggarwal, A. K. Sahoo, S. Bag, V. Kaliginedi, M. Jain, and P. K. Maiti, {Phys. Rev. E} {\bf103}, 032411 (2021).


\bibitem{so} S. O. Kelley, E. M. Boon, J. K. Barton, N. M. Jackson, and M. G. Hill, {Nucleic Acids Res.}  {\bf 27}, 4830 (1999).
\bibitem{bgi} B. Giese and S. Wessely, {Angew. Chem. Int. Ed.} {\bf39}, 3490 (2000).
\bibitem{pkb} P. K. Bhattacharya and J. K. Barton, {J. Am. Chem. Soc.} {\bf 123}, 8649 (2001).
\bibitem{fsh} F. Shao, M. A. O'Neill, and J. K. Barton, {Proc. Natl. Acad. Sci. USA} {\bf 101}, 17914 (2004).
\bibitem{jhi} J. Hihath, B. Xu, P. Zhang, and N. Tao, {Proc. Natl. Acad. Sci. USA} {\bf102}, 16979 (2005).
\bibitem{uos} Y. Osakada, K. Kawai, M. Fujitsuka, and T. Majima, {Nucleic Acids Res.} {\bf 36}, 5562 (2008).
\bibitem{ned} N. Edirisinghe, V. Apalkov, J. Berashevich, and T. Chakraborty, {Nanotechnology} {\bf 21}, 245101 (2010).
\bibitem{mhl} M. H. Lee, S. Avdoshenko, R. Gutierrez, and G. Cuniberti, {Phys. Rev. B} {\bf 82}, 155455 (2010).
\bibitem{smi} S. Mishra, V. S. Poonia, C. Fontanesi, R. Naaman, A. M. Fleming, and C. J. Burrows, {J. Am. Chem. Soc.} {\bf 141}, 123 (2019).


\bibitem{ncse} N. C. Seeman, {Nature} {\bf421}, 427 (2003).
\bibitem{hyan} H. Yan, T. H. LaBean, L. Feng, and J. H. Reif, {Proc. Natl. Acad. Sci. USA} {\bf 100}, 8103 (2003).
\bibitem{rpg1} R. P. Goodman, R. M. Berry, and A. J. Turberfield, {Chem. Commun.}  1372 (2004).
\bibitem{rpg2} R. P. Goodman, I. A. T. Schaap, C. F. Tardin, C. M. Erben, R. M. Berry, C. F. Schmidt, and A. J. Turberfield, {Science} {\bf 310}, 1661 (2005).
\bibitem{pwkr} P. W. K. Rothemund, {Nature} {\bf 440}, 297 (2006).
\bibitem{smdo} S. M. Douglas, H. Dietz, T. Liedl, B. H\"{o}gberg, F. Graf, and W. M. Shih, {Nature} {\bf 459}, 414 (2009).
\bibitem{esan} E. S. Andersen, M. Dong, M. M. Nielsen, K. Jahn, R. Subramani, W. Mamdouh, M. M. Golas, B. Sander, H. Stark, C. L. P. Oliveira, J. S. Pedersen, V. Birkedal, F. Besenbacher, K. V. Gothelf, and J. Kjems, {Nature} {\bf 459}, 73 (2009).


\bibitem{wz} W. Zhu, A.-M. Guo, and Q.-F. Sun, {Front. Phys.} {\bf 9}, 774 (2014).
\bibitem{slit} S. Li, T. Tian, T. Zhang, X. Cai, and Y. Lin, {Materials Today} {\bf 24}, 57 (2019).


\bibitem{asw} A. S. Walsh, H. Yin, C. M. Erben, M. J. A. Wood, and A. J. Turberfield, {ACS Nano} {\bf 5}, 5427 (2011).
\bibitem{zxia} Z. Xia, P. Wang, X. Liu, T. Liu, Y. Yan, J. Yan, J. Zhong, G. Sun, and D. He, {Biochemistry} {\bf 55}, 1326 (2016).
\bibitem{qlid} Q. Li, D. Zhao, X. Shao, S. Lin, X. Xie, M. Liu, W. Ma, S. Shi, and Y. Lin, {ACS Appl. Mater. Interfaces} {\bf 9}, 36695 (2017).
\bibitem{hdi} H. Ding, J. Li, N. Chen, X. Hu, X. Yang, L. Guo, Q. Li, X. Zuo, L. Wang, Y. Ma, and C. Fan, {ACS Cent. Sci.} {\bf 4}, 1344 (2018).


\bibitem{hpei} H. Pei,  N. Lu,  Y. Wen,  S. Song,  Y. Liu,  H. Yan, and  C. Fan, {Adv. Mater.} {\bf 22} 4754 (2010).
\bibitem{nluh} N. Lu, H. Pei, Z. Ge, C. R. Simmons, H. Yan, and C. Fan, {J. Am. Chem. Soc.} {\bf 134}, 13148 (2012).
\bibitem{aabi} A. Abi, M. Lin, H. Pei, C. Fan, E. E. Ferapontova, and X. Zuo, {ACS Appl. Mater. Interfaces} {\bf 6}, 8928 (2014).
\bibitem{clix} C. Li, X. Hu, J. Lu, X. Mao, Y. Xiang, Y. Shu and G. Li, {Chem. Sci.} {\bf9}, 979 (2018).


\bibitem{kjca} K. J. Cash, F. Ricci, and K. W. Plaxco, {J. Am. Chem. Soc.} {\bf 131}, 6955 (2009).
\bibitem{ywen} Y. Wen, H. Pei, Y. Wan, Y. Su, Q. Huang, S. Song, and C. Fan, {Anal. Chem.} {\bf 83}, 7418 (2011).
\bibitem{zgem} Z. Ge, M. Lin, P. Wang, H. Pei, J. Yan, J. Shi, Q. Huang, D. He, C. Fan, and X. Zuo, {Anal. Chem.} {\bf 86}, 2124 (2014).
\bibitem{mli} M. Lin, P. Song, G. Zhou, X. Zuo, A. Aldalbahi, X. Lou, J. Shi, and C. Fan, {Nat. Protoc.} {\bf 11}, 1244 (2016).
\bibitem{pso1} P. Song, M. Li, J. Shen, H. Pei, J. Chao, S. Su, A. Aldalbahi, L. Wang, J. Shi, S. Song, L. Wang, C. Fan, and X. Zuo, {Anal. Chem.} {\bf 88}, 8043 (2016).
\bibitem{pso2} P. Song, J. Shen, D. Ye, B. Dong, F. Wang, H. Pei, J. Wang, J. Shi, L. Wang, W. Xue, Y. Huang, G. Huang, X. Zuo, and C. Fan, {Nat. Commun.} {\bf 11}, 838 (2020).


\bibitem{amg1} A.-M. Guo and Q.-F. Sun, {Phys. Rev. Lett.} {\bf 108}, 218102 (2012).
\bibitem{sd} \textsl{Electronic Transport in Mesoscopic Systems}, edited by S. Datta (Cambridge University Press, Cambridge, England, 1995).
\bibitem{hzx} H. Zhang, X.-Q. Li, P. Han, X. Y. Yu, and Y. Yan, {J. Chem. Phys.} {\bf 117}, 4578 (2002).
\bibitem{kse} K. Senthilkumar, F. C. Grozema, C. F. Guerra, F. M. Bickelhaupt, F. D. Lewis, Y. A. Berlin, M. A. Ratner, and L. D. A. Siebbeles, {J. Am. Chem. Soc.} {\bf 127}, 14894 (2005).
\bibitem{lgd} L. G. D. Hawke, G. Kalosakas, and C. Simserides, {Eur. Phys. J. E} {\bf 32}, 291 (2010).
\bibitem{sro} S. Roche, Phys. Rev. Lett. {\bf91}, 108101 (2003).

\bibitem{yuz} Y. Zhu, C.-C. Kaun, and H. Guo, {Phys. Rev. B} {\bf 69}, 245112 (2004).
\bibitem{gam} A.-M. Guo and Q.-F. Sun, {Phys. Rev. B} {\bf 86}, 115441 (2012).
\bibitem{gam1} L. Chen, F. Ouyang, S. Ma, T.-F. Fang, A.-M. Guo, and Q.-F. Sun, {Phys. Rev. B} {\bf 101}, 115417 (2020).
\bibitem{eli} E. Zerah-Harush and Y. Dubi, {Phys. Rev. Res.} {\bf 2}, 023294 (2020).
\bibitem{pjh} P.-J. Hu, S.-X. Wang, X.-H Gao, Y.-Y. Zhang, T.-F. Fang, A.-M. Guo, and Q.-F. Sun, {Phys. Rev. B} {\bf 102}, 195406 (2020).

\end{references}
\end{document}